\documentclass[aps,prstab,superscriptaddress,floatfix,twocolumn,amsfonts,amsmath,amssymb,amsart]{revtex4-1}
\usepackage{tikz-feynman}
\usepackage{comment}
\usepackage{graphicx} 
\usepackage{float}
\usepackage{dcolumn} 
\usepackage{bm} 
\usepackage[colorlinks, breaklinks=true,linkcolor=red, citecolor=blue, linktocpage=true]{hyperref}
\usepackage{subcaption}
\renewcommand{\vec}{\mathbf}
\newcommand{\RE}{\mathrm{Re}}
\newcommand{\IM}{\mathrm{Im}}
\begin{document}

\title{Theory of Band Gap Reduction Due to Conduction Electrons in 2D TMDs: Imaginary Frequency Formalism}
\author{Jack N. Engdahl}
\thanks{j.engdahl@student.unsw.edu.au}
\affiliation{School of Physics, University of New South Wales, Sydney 2052, Australia}
\author{Harley D. Scammell}
\affiliation{School of Mathematical and Physical Sciences, University of Technology Sydney, Ultimo, NSW 2007, Australia}
\author{Dmitry K. Efimkin}
\affiliation{School of Physics and Astronomy, Monash University, Victoria 3800, Australia}
\author{Oleg P. Sushkov}
\affiliation{School of Physics, University of New South Wales, Sydney 2052, Australia}
\date{\today}

\begin{abstract}
   {Two Dimensional (2D) Transition Metal Dichalcogenides (TMDs) possess a large direct band gap which has been experimentally observed to shrink with increasing charge carrier density (doping). The effect has been the subject of theoretical study in recent years using various approaches and approximations. In this work we develop the theory of bandgap renormalization based on Feynman diagrammatic technique in the imaginary frequency formalism. We consider dynamical screening from conduction band electrons using the random phase approximation (RPA), as well as screening from a metallic gate. While our theory is general for any 2D semiconductor, to be specific we consider MoS$_2$ and WSe$_2$ and compare with available experimental data. In both cases we calculate large band gap renormalization that reaches several hundred meV at relatively low carrier density. This is in good agreement with experimental data.}
\end{abstract}

\maketitle
\section{Introduction}
Atomically thin Transition Metal Dichalcogenides (TMDs) are a class of two dimensional (2D) semiconducting materials with quadratic band structure and large direct band gaps at the K point of the Brillouin zone~\cite{wang_colloquium_2018,scharf_dynamical_2019}.  The conduction and valence bands are spin split by the spin-orbit interaction, with this splitting much greater in the valence bands due to particle hole asymmetry,
and the spin and valley degrees of freedom are coupled~\cite{kormanyos_k_2015,stier_exciton_2016,rostami_effective_2013,wang_colloquium_2018,scharf_dynamical_2019}.

The electronic properties of 2D TMDs make them an ideal platform for the study of exciton physics \cite{chen_luminescent_2019,wu_exciton_2015,liu_hybrid_2021,goldstein_ground_2020,liu_landau-quantized_2020,bange_ultrafast_2023,zhu_-plane_2023}.
Because of large binding energies excitons dominate the absorption spectrum even at room temperature~\cite{wang_colloquium_2018,scharf_dynamical_2019, chen_luminescent_2019, meckbach_influence_2018,qiu_screening_2016,qiu_giant_2019,bera_atomlike_2021,chernikov_population_2015,pogna_photo-induced_2016,cunningham_photoinduced_2017,yao_optically_2017} .  In terms of applications, due to their direct band gap in the optical range 2D TMDs are ideal for the creation of optoelectronic devices~\cite{wang_electronics_2012,mueller_exciton_2018,taffelli_mos2_2021}, including the fabrication of heterostructure stacks which are an excellent environment for light-matter coupling in the form of exciton-polaritons~\cite{dufferwiel_valley_2018,zhao_exciton_2023,zhang_exciton_2020}, or in the development of monolayer TMD based photodetectors~\cite{yin_single-layer_2012,velusamy_flexible_2015,chang_monolayer_2014,zhang_highgain_2013,lopez-sanchez_ultrasensitive_2013,gonzalez_marin_mos2_2019}.

Whether it is for the study of exciton physics or for device applications, it is imperative to accurately know the intrinsic band gap of the 2D TMD, which may be measured experimentally through ARPES \cite{lee_time-resolved_2021,kang_universal_2017,nguyen_visualizing_2019}, STS \cite{zhang_bandgap_2016,qiu_giant_2019} or differential reflectance spectroscopy \cite{qiu_giant_2019}. Furthermore, one must also understand the effect of many body screening on the Coulomb interaction in the system, which not only effects the exciton spectrum~\cite{qiu_screening_2016,qiu_giant_2019,gao_dynamical_2016,karmakar_observation_2021,steinhoff_frequency-dependent_2018,cunningham_photoinduced_2017,pogna_photo-induced_2016,chernikov_population_2015,yao_optically_2017}, but also has a significant effect on bandgap~\cite{gao_dynamical_2016,liang_carrier_2015,qiu_giant_2019,qiu_screening_2016,zibouche_gw_2021,liu_direct_2019,kang_universal_2017,steinhoff_frequency-dependent_2018,ataei_competitive_2021,fernandez_renormalization_2020,meckbach_giant_2018,yao_optically_2017,cunningham_photoinduced_2017,pogna_photo-induced_2016,bera_atomlike_2021,tian_broadband_2022,wood_evidence_2020,chernikov_population_2015,nguyen_visualizing_2019}.

The addition of electrons in the conduction band causes the band gap to
  decrease significantly compared to its value in the insulator. The physics behind the gap reduction is conceptually straightforward.
  Let ${\cal D}^{(0)}$ be the band gap in a hypothetical band insulator without
  electron-electron Coulomb interaction. Switching the interaction on increases
  the gap, ${\cal D}^{(ins)} > {\cal D}^{(0)}$, where ${\cal D}^{(ins)}$ is the gap
    in the
  real band insulator. This effect is similar to the electromagnetic
  contribution to the mass of electron in
  Quantum Electrodynamics (QED)~\cite{beresteckij_quantum_2008}.
  Conduction band electrons screen the Coulomb interaction, therefore the addition of electrons in the conduction band leads to the reduction
  of the Coulomb interaction and hence reduces the insulating band gap towards
  ${\cal D}^{(0)}$. 
  We will show that at a small density of conduction electrons $n$ the gap
  reduction is ${\cal D} ={\cal D}^{(ins)}- A n^{1/3}$, where $A$ is a constant that depends on microscopic details.
  Because of the small power, $1/3$, the reduction is significant even at
  very low $n$.

  There are several different approximations used in the literature
  to account for the screening and to calculate the gap reduction.
  These methods include simple static approximation
  to different kinds of plasmon pole approximations~\cite{gao_dynamical_2016,gao_renormalization_2017,liang_carrier_2015,steinhoff_frequency-dependent_2018}.
  The different methods give qualitatively similar results, but quantitatively
  the results are quite different.
  Previously, theoretical calculation using the full RPA summation
  have been performed  in Ref.~\cite{faridi_quasiparticle_2021}.
  This calculation has been done in the real frequency domain. In the present work,
  following the QED techniques \cite{beresteckij_quantum_2008},  we develop the method of RPA analysis of the band gap
  renormalization working in the imaginary frequency domain. The method leads to
  significantly simpler calculations compared to that in the real frequency
  domain. We compare our theory to available experimental data, finding that our theory fits the data very well, and also compare to approximate calculations such as a plasmon pole approximation.
  We have also shown that the gate screening, which is usually disregarded, is significant for realistic devices.

  The structure of the paper is as follows. In Sec.~\ref{sec:gate} we present the self energy for an insulator with metallic gate and derive the expression for correction to the insulator gap due to gate screening. Following from this, in Sec.~\ref{sec:self_energy} we consider the self energy in the metallic state and derive an expression for correction to the band gap due to screening from both conduction electrons and a metallic gate, where we work in terms of imaginary frequency. This expression is general, but to present results we must choose a specific system. Thus in Sec.~\ref{sec:Pol} we discuss the RPA polarization for 2D TMDs and in Sec.~\ref{sec:gap_theory} we investigate the band gap renormalization in the low density regime to explain the very sharp dependence on density.
We compare our theory to experimental results in Sec.~\ref{sec:experiment} and find that they are in good agreement. We summarise our key findings in Sec.~\ref{sec:summary}. Finally in Appendix~\ref{sec:compare} we compare our RPA results with models often used in simplified calculations, the static screening model and the plasmon pole model.


\section{The gap reduction in the insulator due to the gate} \label{sec:gate}
\begin{figure} [t]
  \includegraphics[height=0.42\linewidth]{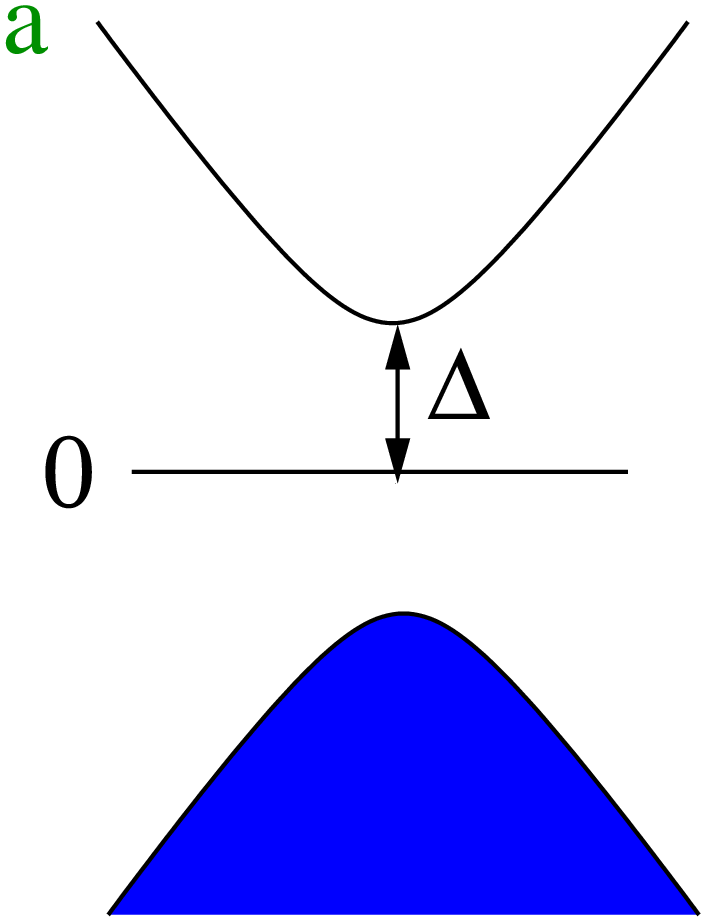}
  \hspace{40pt}
\includegraphics[height=0.30\linewidth]{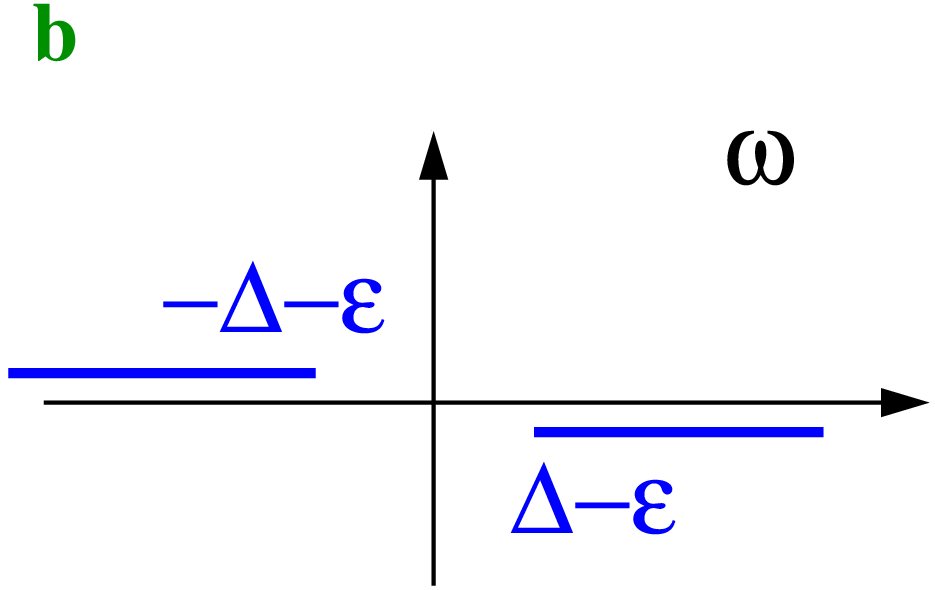}
\vspace{5pt}
\caption{Panel a: filling of electron states in the insulator.
  Panel b: singularities of the integrand in Eq.(\ref{se1}),
  the interaction ${\cal V}^{(ins)}$ has no singularities
  and the blue lines show cuts of the Green's function $G^{(ins)}(\epsilon+\omega)$.
}
\label{D1}
\end{figure}

To illustrate our method we begin with the slightly simplified scenario of band gap renormalization of an insulator due to screening from a metallic gate. We start from the electron self energy using the Feynman method.
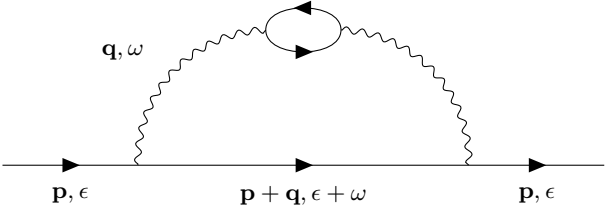
\begin{figure}
	\centering
	\begin{tikzpicture} 
		\begin{feynman}
			\vertex at (0.9, -0.4) {\(\vec{p}, \epsilon\)};
			\vertex at (7.1, -0.4) {\(\vec{p}, \epsilon\)};
			\vertex at (0, 0) (a) ;
			\vertex at (1.8, 0) (b) ;
			\vertex at (4, -0.4)  {\(\vec{p}+\vec{q}, \epsilon+\omega\)};
			\vertex at (6.2, 0) (c);
			\vertex at (8, 0) (d) ;
			\vertex at (3.5, 1.8) (e);
			\vertex at (1.6, 1.5) {\(\vec{q}, \omega\)};
			\vertex at (4.5, 1.8) (f);
			\diagram*{
				(a) -- [fermion] (b) -- [fermion] (c) --[fermion] (d),
				(b) -- [boson, quarter left, looseness=1.0] (e) -- [fermion, half right, looseness=1.0] (f) -- [fermion, half right, looseness=1.0] (e), (f) -- [boson, quarter left, looseness=1.0] (c)
			};
		\end{feynman}
	\end{tikzpicture}
	\caption{Electron self energy diagram, straight line corresponds to the electron
	  and the wiggly line to the Coulomb interaction. The bubble corresponds to the RPA polarization operator. While we show only one bubble, the entire RPA chain is accounted for in this procedure.
        }
	\label{SEf}
\end{figure}
The self energy is given by diagram in Fig.~\ref{SEf}, where $\epsilon$, $\vec{p}$ are the external frequency and momentum and $\omega$, $\vec{q}$ are the frequency and momentum of interaction. The
          corresponding analytic formula is         
	  \begin{eqnarray}
            \label{se1}
	    \Sigma_p(\epsilon) = -\lim_{\tau\to0^+} \sum_{\bf q}\int
            e^{i\omega\tau}G_{\vec{p}+\vec{q}}(\epsilon+\omega)
            {\cal V}_q(\omega)\frac{d\omega}{2\pi i}.
        	\end{eqnarray}
          Here $G$ is the electron Green's function and ${\cal V}_q(\omega)$ is
          the RPA Coulomb interaction.
First consider the insulator with half band gap $\Delta$ shown in Fig.~\ref{D1}a, 
with the electron Green's function
        \begin{eqnarray}
          \label{g0}
          G^{(ins)}_p(\epsilon)&=&
          \frac{{\cal P}_-
          }{\epsilon+\left(\Delta+\epsilon_p^{(h)}\right)-i0}+
          \frac{{\cal P}_+}{\epsilon-\left(\Delta+\epsilon_p^{(e)}\right)+i0}
          \nonumber\\
          \epsilon_p^{(e)}&=&  \frac{p^2}{2m_e}\ , \ \ \ \
          \epsilon_p^{(h)}= \frac{p^2}{2m_h}\ .
          \end{eqnarray}
        The RPA interaction in this case is the instantaneous Coulomb
        interaction
        \begin{eqnarray}
          \label{Vins}
{\cal V}^{(ins)}_q={\cal I} \ V^{(ins)}_q\ , \ \ \ \ 
V^{(ins)}_q= \frac{2\pi e^2}{\varepsilon_q q/(1-e^{-2qd})}
        \end{eqnarray}
        Here besided the usual Coulomb repulsion we account three
        additional points specific for TMD devices. (i) The polarization operator of the zero temperature insulator is practically zero. (ii) Due to the 2D geometry of the system the interaction is described by a Rytova-Keldysh potential, which accounts for differences in the dielectric constant between layers of the system more accurately than the standard 2D Coloumb potential. In this regime there is significant
        momentum
        dependence of the effective dielectric constant,
        $\varepsilon \to \varepsilon_q=\varepsilon (1+r_0q)$,
        Ref.~\cite{cudazzo_dielectric_2011}. (iii) A metallic gate at distance $d$ from  TMD results in
        electrostatic screening.        
Note that both the Green's function and the interaction are $2\times 2$ matrices where index $i=1$
corresponds to the positive energy states and $i=2$ corresponds to the negative energy states.
The projectors and the unity matrix are
\begin{eqnarray}
  {\cal P}_+=          \left(
          \begin{array}{cc}
            1& 0\\
            0& 0
          \end{array}
          \right) \ \ \ \
            {\cal P}_-=          \left(
          \begin{array}{cc}
            0& 0\\
            0& 1
          \end{array}
          \right) \ \ \ \
  {\cal I}=          \left(
          \begin{array}{cc}
            1& 0\\
            0& 1
          \end{array}
          \right)
\end{eqnarray}          
The Green's function is a diagonal matrix,
the interaction in principle has off-diagonal terms, but in TMDs because of the very large band gap
 ${\cal D}=\Delta\sim2$ eV the
wave function overlap between electrons in the conduction and valence bands is vanishingly small and the interaction matrix is thus approximately diagonal.
Hence, the electron self energy in the insulator reads
	  \begin{eqnarray}
            \label{seins}
      \Sigma^{(ins)}_p(\epsilon) = -\lim_{\tau\to0^+} \sum_{\bf q}\int e^{i\omega\tau}
G^{(ins)}_{\vec{p}+\vec{q}}(\epsilon+\omega){\cal V}^{(ins)}_q\frac{d\omega}{2\pi i}
        	\end{eqnarray}
Cuts  of this equation integrand  in the complex $\omega$-plane
are shown in Fig.~\ref{D1}b. They are determined by the analytical properties of the Green's function
and hence  start from $\Delta-\epsilon$ and $-\Delta-\epsilon$.

\begin{figure} [t]
	\centering
	\vspace{-20pt}
	\includegraphics[width=0.9\linewidth]{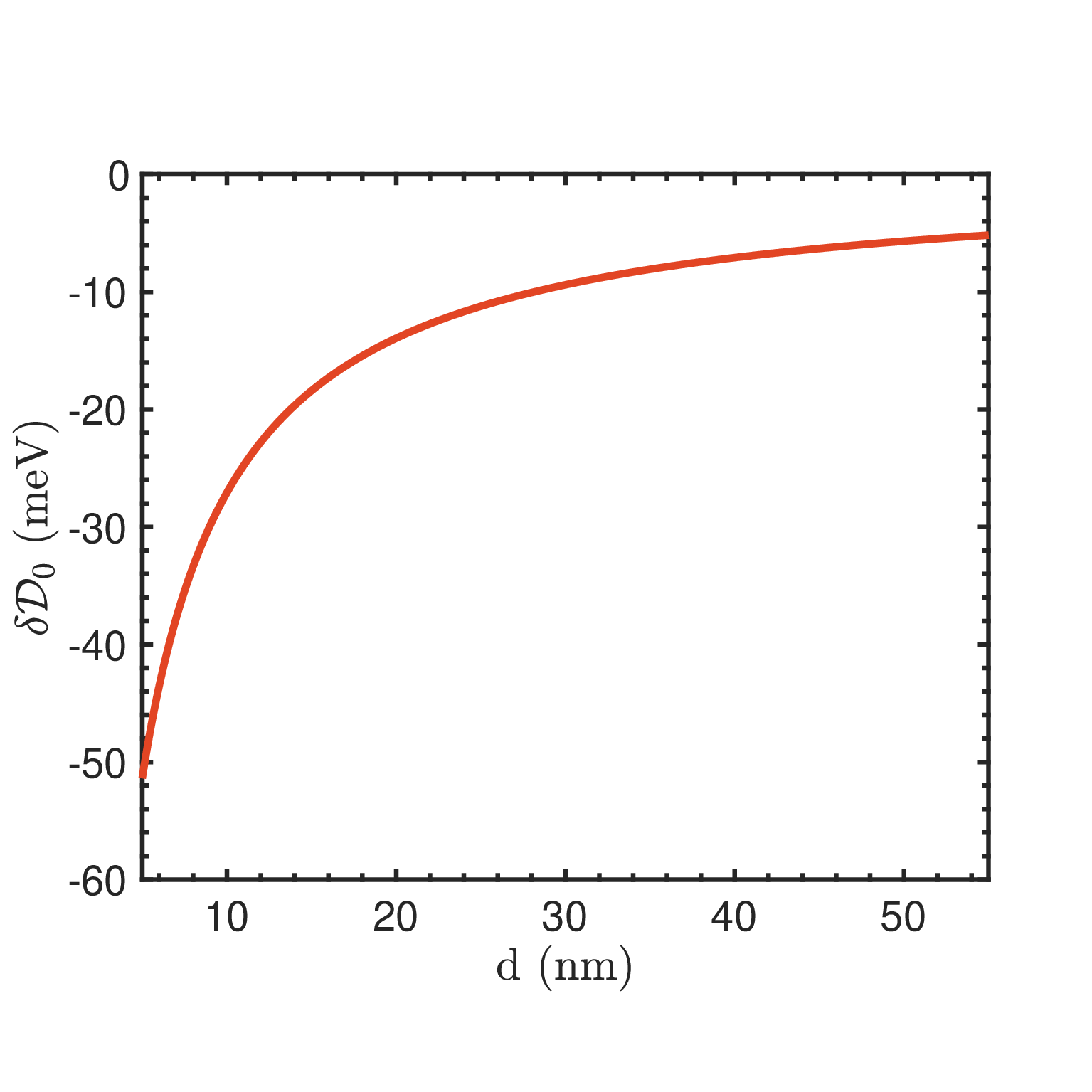}
	\vspace{-15pt}
	\caption{Insulator band gap renormalization due to gate screening. The device is MoS$_2$ on a SiO$_2$ substrate.}
	\label{fig:gate}
\end{figure}

The band gap variation due to the interaction is
\begin{eqnarray}
  \label{gv1}
\delta{\cal D}_0=  \Sigma^{(ins+)}_{p=0}(\epsilon=\Delta)-
\Sigma^{(ins-)}_{p=0}(\epsilon=-\Delta), 
\end{eqnarray}
where superscript $ins+$ ($ins-$) refers to the conduction (valence) band in the insulator.
Only the second term in Eq.\ref{gv1} is nonzero. We calculate the
gap variation due to finite $d$ (relative to $d\to\infty$)
\begin{eqnarray}
\label{igr}
  \delta{\cal D}_0 &=&  \lim_{\tau \to 0^+} \int e^{i\omega \tau}
  \frac{V_q^{(ins)}(d)-V_{q}^{(ins)}(d\to\infty)}{\omega + \epsilon_q^{(h)} -i0}
  \nonumber\\
  &\times&  \frac{d\omega}{2 \pi i} \frac{d^2q}{(2 \pi)^2}
  =-\frac{e^2}{\varepsilon}\int_0^\infty \frac{e^{-2qd}}{1+r_0q}dq\nonumber\\
  &\approx& \frac{e^2}{2\varepsilon d}\left(1-\frac{r_0}{2d}+ . . .\right)
\end{eqnarray}
Let us take MoS$_2$ on SiO$_2$ substrate as an example.
For this case we use the following parameters, $\varepsilon = 2.5$ and
$r_0 = 3.4$ nm/$\varepsilon$, see Refs.~\cite{wu_exciton_2015,zhang_absorption_2014}.  We plot band gap renormalization due to gate screening for this device in Fig.~\ref{fig:gate}. 

Making speial note of $d=10$ nm, which is typical of such devices,
we find gate induced band gap reduction $\delta{\cal D}_0\approx -27meV$. This effect is non-negligible, but the more significant effect is due to conduction electrons directly in the plane of the 2D TMD. We consider this in the following section.

\section{Electron Self Energy with conduction electrons,
  Wick Rotation to Imaginary Axis, the band gap variation} \label{sec:self_energy}

\begin{figure} [t]
  \includegraphics[height=0.42\linewidth]{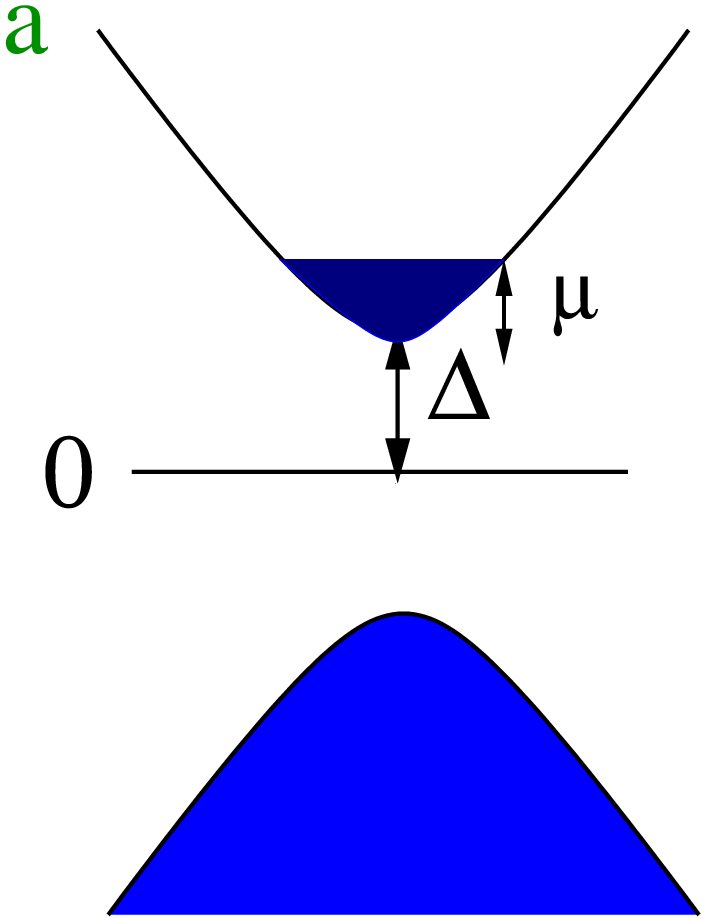}
  \hspace{35pt}
\includegraphics[height=0.30\linewidth]{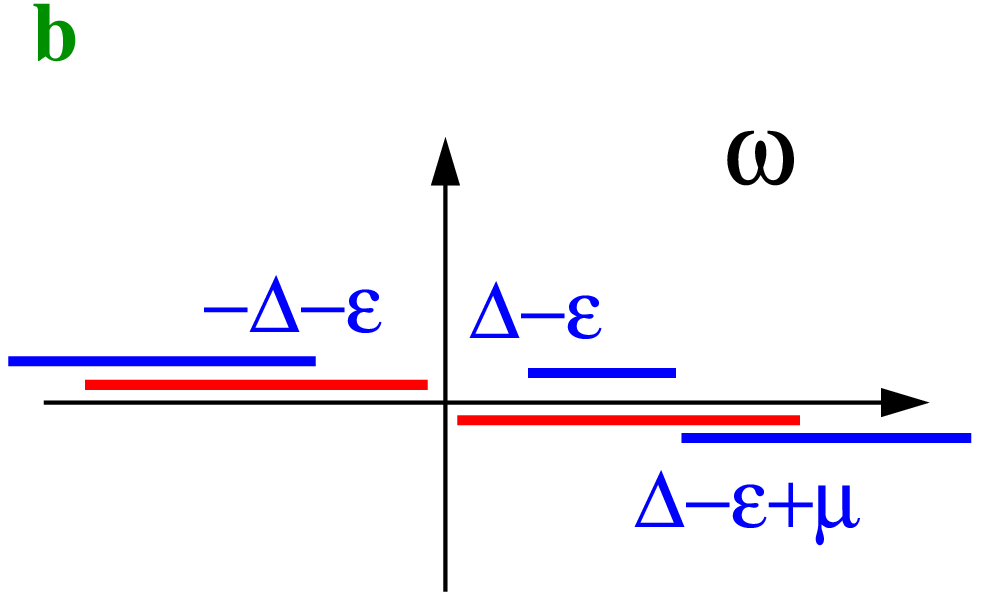}
\vspace{5pt}
\caption{Panel a: filling of electron states in the metal.
  Panel b: singularities of the integrand in Eq.(\ref{se1}), the red lines
  show cuts of the interaction ${\cal V}^{(m)}(\omega)$ and the blue lines show
  cuts of the Green's function $G^{(m)}(\epsilon+\omega)$.
}
\label{D2}
\end{figure}
Consider now the metallic regime, where conduction electrons fill the conduction band up to the Fermi level $\mu$. Filling of states in a metal is shown in Fig.~\ref{D2}a.
        The RPA interaction in this case is
        \begin{eqnarray}
          \label{Vm}
    {\cal V}^{(m)}_q(\omega)&=&{\cal I} \ V^{(m)}_q(\omega)\nonumber\\ 
                V^{(m)}_q(\omega)&=&
        \frac{2\pi e^2}{\varepsilon_qq/(1-e^{-2qd})- 2\pi e^2\Pi_q(\omega)}
\end{eqnarray}
where $\Pi_q(\omega)$ is the polarization operator in the metallic state.
  In the metallic state cuts of the integrand of Eq.(\ref{se1}) in the complex $\omega$-plane  are
        shown in Fig.~\ref{D2}b.
The cuts of the interaction start from $\pm 0$, and a part of the Green's
function cut  is shifted to the upper half plane due to conduction electrons. This prevents a simple procedure of Wick's rotation to the imaginary axis as the contour of integration would now cross a branch cut during rotation. 
To resolve this complication we employ the Sokhotski–Plemelj theorem to represent the Green's function as
        \begin{align}
          \label{g1}
        G^{(m)}_p(\epsilon)&=\frac{ n_p^{(v)}{\cal P}_-}{\epsilon+\left(\Delta+\epsilon_p^{(h)}\right)-i0}+
          \frac{\left(1- n_p^{(v)}\right){\cal P}_-}{\epsilon+\left(\Delta+\epsilon_p^{(h)}\right)+i0} 
          \nonumber\\&+ \frac{ n_p^{(c)}{\cal P}_+}{\epsilon-\left(\Delta+\epsilon_p^{(e)}\right)-i0}+
          \frac{\left(1- n_p^{(c)}\right){\cal P}_+}{\epsilon-\left(\Delta+\epsilon_p^{(e)}\right)+i0}  \nonumber\\
          &=G^{(ins)}_p(\epsilon)+ {\cal P}_+
          2\pi i n_p^{(c)}\delta\left(\epsilon-\Delta-\epsilon_p^{(e)} \right),
          \end{align}
where $n_p^{(c)}$ $\left(n_p^{(v)}\right )$is the electron occupation number in the conduction (valence) band.
Hence the self energy in the metallic state (with $n_p^{(v)}=1$) can be written as
\begin{eqnarray}
\label{sem}
\Sigma^{(m)}_p(\epsilon) &=& -\lim_{\tau\to0^+} \sum_{\bf q}\int e^{i\omega\tau}
{\cal V}^{(m)}_q(\omega)
\left[G^{(ins)}_{\vec{p}+\vec{q}} (\epsilon+\omega)\right.\nonumber\\
  &+&\left.{\cal P}_+2\pi i n_{\vec{p}+\vec{q}}^{(c)}
  \delta\left(\epsilon+\omega-\Delta-\epsilon_{\vec{p}+\vec{q}}^{(e)}\right)
  \right]\frac{d\omega}{2\pi i}\nonumber\\
&=&
-\lim_{\tau\to0^+} \sum_{\bf q}\int e^{i\omega\tau}{\cal V}^{(m)}_q(\omega)
G^{(ins)}_{\vec{p}+\vec{q}}(\epsilon+\omega)\frac{d\omega}{2\pi i}\nonumber\\
&-&{\cal P}_+\sum_{\bf q} n_{\vec{p}+\vec{q}}^{(c)}
V^{(m)}_q\left(\Delta+\epsilon_{\vec{p}+\vec{q}}^{(e)}-\epsilon\right)
	\end{eqnarray}
The renormalized self energy operator is the difference between
metal, Eq.(\ref{sem}), and insulator, Eq.(\ref{seins}),
$\Sigma^{(R)}=\Sigma^{(m)}-\Sigma^{{ins}}$,
\begin{eqnarray}
  \label{ser}
  \Sigma^{(R)}_p(\epsilon) &=& -\lim_{\tau\to0^+} \sum_{\bf q}\int
  e^{i\omega\tau}\left[{\cal V}^{(m)}_q(\omega)-
    {\cal V}^{(ins)}_q\right]\nonumber\\
&\times&  G^{(ins)}_{\vec{p}+\vec{q}}  (\epsilon+\omega)
\frac{d\omega}{2\pi i}\nonumber\\
&-&{\cal P}_+\sum_{\bf q} n_{\vec{p}+\vec{q}}^{(c)}
V^{(m)}_q\left(\Delta+\epsilon_{\vec{p}+\vec{q}}^{(e)}-\epsilon\right)
\end{eqnarray}
We stress that we compare the metal and the insulator with the same
gate distance $d$. Therefore, the difference
${\cal V}^{(m)}_q(\omega)-{\cal V}^{(ins)}_q \to 0$ at large
frequencies, $\omega \to \infty$.
Hence, the regularizor $e^{i\omega\tau}$ can be omitted in Eq.(\ref{ser}).
After that, having in mind that $-\Delta \leq \epsilon \leq \Delta$, the Wick
rotation, $\omega \to i\xi$  can be performed in the 1st term of (\ref{ser}).
Hence, we arrive at
\begin{eqnarray}
  \label{ser1}
  \Sigma^{(R)}_p(\epsilon) &=& -\sum_{\bf q}\int\left[{\cal V}^{(m)}_q(i\xi)-
    {\cal V}^{(ins)}_q\right]G^{(ins)}_{\vec{p}+\vec{q}} (\epsilon+i\xi)
  \frac{d\xi}{2\pi}\nonumber\\
  &-&{\cal P}_+\sum_{\bf q} n_{\vec{p}+\vec{q}}^{(c)}
  V^{(m)}_q\left(\Delta+\epsilon_{\vec{p}+\vec{q}}^{(e)}  -\epsilon\right)
\end{eqnarray}
Similar to Eq.(\ref{gv1}) the band gap variation in the metal compared to that in
the insulator is
$\delta{\cal D}=   \Sigma^{(R+)}_{p=0}(\epsilon=\Delta)- \Sigma^{(R-)}_{p=0}(\epsilon=-\Delta)$,
hence
\begin{eqnarray}
  \label{gapv}
  \delta{\cal D}&=&\sum_{\bf q}\int \left[V^{(m)}_q(i\xi)-
    V^{(ins)}_q\right]\nonumber\\
  &\times&  \left[\frac{\epsilon_q^{(e)}}{\xi^2+(\epsilon_q^{(e)})^2}
+\frac{\epsilon_q^{(h)}}{\xi^2+(\epsilon_q^{(h)})^2}
    \right]\frac{d\xi}{2\pi}\nonumber\\
  &-&\sum_{\bf q} n_{\bf q}^{(c)}V^{(m)}_q(\epsilon_q^{(e)})
\end{eqnarray}
Note that the second term in this equation has a small  imaginary part related to the
lifetime of a hole at the bottom of the conduction band. 
The variation of the physical band gap is given by the real part of Eq.~(\ref{gapv}).

\section{Polarization Operator of electrons} \label{sec:Pol}
\begin{figure}
	\centering
	\begin{tikzpicture} 
		\begin{feynman}
			\vertex at (2.8, 1.3)  {\(\vec{q}, \omega\)};
			\vertex at (2.2, 1.0) (a);
			\vertex at (3.6, 1.0) (b);
			\vertex at (5.4, 1.0) (c);
			\vertex at (6.8, 1.0) (d);
			\diagram*{
				(a) -- [boson] (b) -- [fermion, half right, looseness=1.0] (c) -- [fermion, half right, looseness=1.0] (b), (c) -- [boson] (d)
			};
		\end{feynman}
	\end{tikzpicture}
	\caption{Polarization operator bubble.}
	\label{fig:pol}
\end{figure}
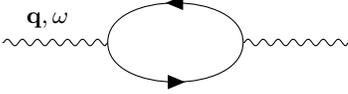

In 2D TMDs the band gap is on the order of ${\cal D} \sim 2$eV.
The Fermi energy of electrons is small compared to the gap,
$\epsilon_F=\mu \ll {\cal D}$.
The polarization operator is entirely due to conduction electrons and is given
by the diagram Fig.~\ref{fig:pol}.

The 2D Feynman polarization operator for parabolic dispersion is
well known~\cite{milstein_effective_2008}. In the real frequency formalism it is  
\begin{eqnarray}
  \label{popre}
&&  \RE \Pi_q(\omega) = -\frac{m_e}{\pi}
  \left[1-\frac{\epsilon_q-\omega}{|\epsilon_q-\omega|}
    {\overline \Theta}(\beta_-) -\frac{\epsilon_q+\omega}{|\epsilon_q+\omega|}
 {\overline \Theta}(\beta_+)\right]\nonumber\\
&&\IM \Pi_q(\omega) = -sgn(\omega)
 \frac{m_e}{\pi}
 \left[{\overline \Theta}(-\beta_-)-{\overline \Theta}(-\beta_+)\right]
 \nonumber\\
&&{\overline \Theta}(\beta)=\sqrt{\beta}\Theta(\beta)\nonumber\\
&& \beta_\pm = \frac{1}{4}\left(1\pm \frac{\omega}{\epsilon_q}\right)^2-
 \frac{\epsilon_F}{\epsilon_q}.
 \end{eqnarray}
Here $\Theta(x)$ is the Heaviside step function defined as $\Theta(x)=1$ for $x\geq0$, $\Theta(x)=0$ for $x<0$; the electron dispersion is
$\epsilon_q=\epsilon_q^{(e)}=\frac{q^2}{2m_e}$.
For the imaginary frequency the polarization operator is
\begin{equation}
\label{pi_xi}
\Pi_q(i\xi) = \frac{-m_e}{\pi} \left ( 1 - 2 \RE \sqrt{
\frac{1}{4}\left(1+ \frac{i\xi}{\epsilon_q}\right)^2-
 \frac{\epsilon_F}{\epsilon_q}}\right).
\end{equation}
The polarization operators (\ref{popre}),(\ref{pi_xi}) correspond to two
  electrons per filled momentum state. 
2D TMDs have two valleys, but each valley is spin split. This coupling of spin and valley degrees of freedom is well known and is due to strong spin orbit coupling, see Refs.~\cite{kormanyos_k_2015,stier_exciton_2016,rostami_effective_2013,wang_colloquium_2018,scharf_dynamical_2019}.
Therefore, there are two different valley/spin states at a given momentum.

\section{The gap reduction due to conduction electrons} \label{sec:gap_theory}
It is important to qualitatively understand the parametric behaviour of the band gap reduction as a function of conduction electron density. There are three parameters of dimension length in Eq.(\ref{gapv}), namely the Bohr radius
   $a_B=\frac{\hbar^2\varepsilon}{ e^2 m_e}$, the dielectric constant momentum dependence scale $r_0$
and the distance to the metallic gate $d$. They are ordered as $a_B < r_0 < d$.
With $\hbar=1$, let $p_F=\sqrt{2\pi n}$ be the Fermi momentum of conduction electrons.
The most important dimensionless parameter is $a_Bp_F$, which is $a_Bp_F\sim  10^{-2}$ at $n=10^{11}$cm$^{-2}$ and $a_Bp_F\sim 1$ at $n=10^{14}$cm$^{-2}$.
Let us first consider the infinitely far gate case,  $d\to\infty$.
Analysis of the imaginary frequency integral in (\ref{gapv}) shows that it is convergent
at $q \sim \frac{p_F}{(a_Bp_F)^{1/3}} \gg p_F$.
At $\frac{p_Fr_0}{(a_Bp_F)^{1/3}}\ll 1$ the length scale $r_0$ may be discarded, which allows for a simple approximate formula in the low
density regime $n \ll \frac{a_B}{2\pi r_0^3} \sim  10^{12}$cm$^{-2}$. Although it should be noted that as the small parameter scales with $n^{1/3}$ the low density regime is significantly lower than $10^{12}$cm$^{-2}$.
In this regime evaluation of integrals in (\ref{gapv}) gives
\begin{eqnarray}
    \label{gld}
    \delta{\cal D}=-1.4\frac{e^2}{\varepsilon a_B}(a_Bp_F)^{2/3}-\frac{5}{12}\frac{p_F^2}{2m_e}.
\end{eqnarray}    
Here the first term comes from the first term in (\ref{gapv})
(imaginary frequency integration) and the second term~\footnote{Note that in taking the limit  $a_B p_F \to 0$ the discarding of small terms linear or higher order in $a_B p_F$ leads to the cancellation of $e^2$ terms in the second term of (\ref{gld}), seemingly implying the existence of band gap renormalizaiton in the non-interacting scenario. However, it is clear by inspection of the interaction (\ref{Vm}) and Eq.~(\ref{gapv}) that formally in the limit $e^2\to0$ this term vanishes as required. }  comes from
the second term in (\ref{gapv}). Clearly, the first term dominates at low density. While the imaginary part in the second term in Eq. (\ref{gapv}) is unrelated to the band gap, still, it is instructive to evaluate it.
The imaginary part is $+\frac{1}{6}\frac{p_F^2}{2m_e}$.

\begin{figure} [t]
	\centering
		\vspace{-20pt}
	\includegraphics[width=0.9\linewidth]{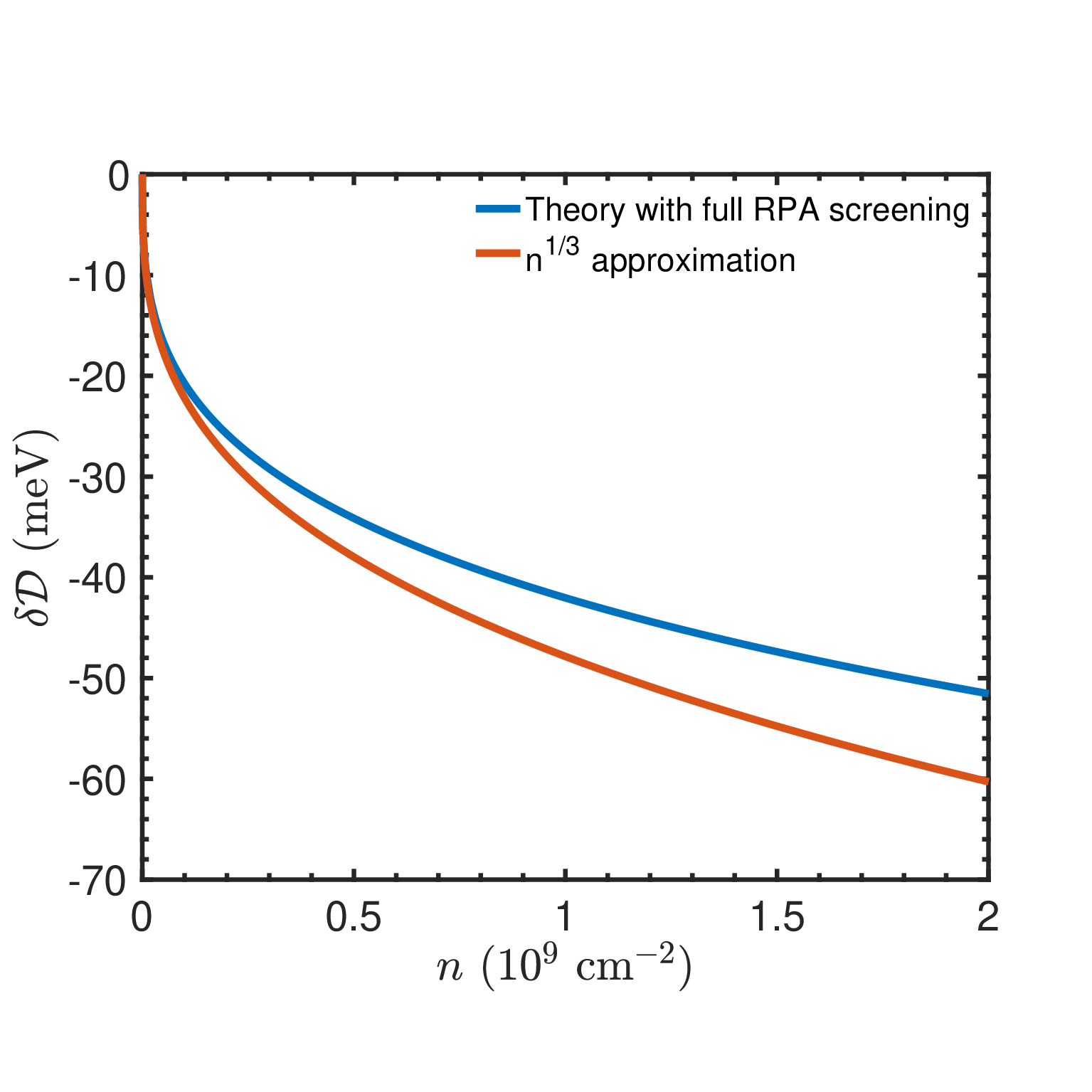}
		\vspace{-15pt}
	\caption{Band gap reduction calculation from Eq. (blue line) compared with the low density approximation Eq. (red line). The device is MoS$_2$ on SiO$_2$ substrate with a gate located at $d\to\infty$.}
	\label{fig:approximation}
\end{figure}

Again, let us take  MoS$_2$ on SiO$_2$ substrate as an example.
The relevant parameters are given in Table.~\ref{table:1}.
\begin{table}[bp]
	\caption{Material parameters for the MoS$_2$ device.}
	\label{table:1}
	\begin{tabular}{lll}
		\hline
		\multicolumn{3}{l}{MoS$_2$. Strata: SiO$_2$, air} \\ \hline
		Parameter & Value & Description  \\ \hline
		$m_e$ & $0.44$ m$_0$ \cite{kormanyos_k_2015} & Effective electron mass \\
		$m_h$ & $0.54$ m$_0$ \cite{kormanyos_k_2015} & Effective hole mass \\
		$\varepsilon$ & $2.5$ \cite{wu_exciton_2015} & Effective dielectric constant \\
		$r_0$ & $3.4$ nm/$\varepsilon$ \cite{wu_exciton_2015,zhang_absorption_2014} & Characteristic screening length 
		\\ \hline
	\end{tabular}
\end{table}
Plots of $\delta{\cal D}$ calculated with the approximation (\ref{gld})
and with exact numerical integration of Eq. (\ref{gapv}) with the above set of
parameters and $d\to\infty$ are presented in Fig.{\ref{fig:approximation}, the plots are identical at $ n \to 0$.
In Fig.~\ref{fig:MoS2_d} we present plots of the band gap reduction versus $n$ 
  with exact numerical integration of Eq. (\ref{gapv}) using the above set of
  parameters for several values of $d$. This figure clearly demonstrates importance of the gate, which significantly effects $\delta{\cal D}$ at low $d$.
  We stress that Fig.~\ref{fig:MoS2_d} shows the band gap reduction of the metal compared to the
  insulator in a set up with the same gate. When comparing two devices with different gate distances the insulator gap reduction relative to the case $d\to\infty$, given by Eq.(\ref{igr}), must be added on top of this, such that
  
  \begin{eqnarray}
  	{\cal D} = {\cal D}_0+\delta{\cal D}_0+\delta{\cal D} \ .
  \end{eqnarray}
  We remind the reader that ${\cal D}_0$ is the gap in the insulator with $d=\infty$,
  ${\cal D}_0+\delta{\cal D}_0$ is the gap in the insulator with finite $d$,
  and $\delta{\cal D}$ is the correction due to conduction electrons
  calculated at a given $d$.

\begin{figure} [t]
	\centering
		\vspace{-20pt}
	\includegraphics[width=0.9\linewidth]{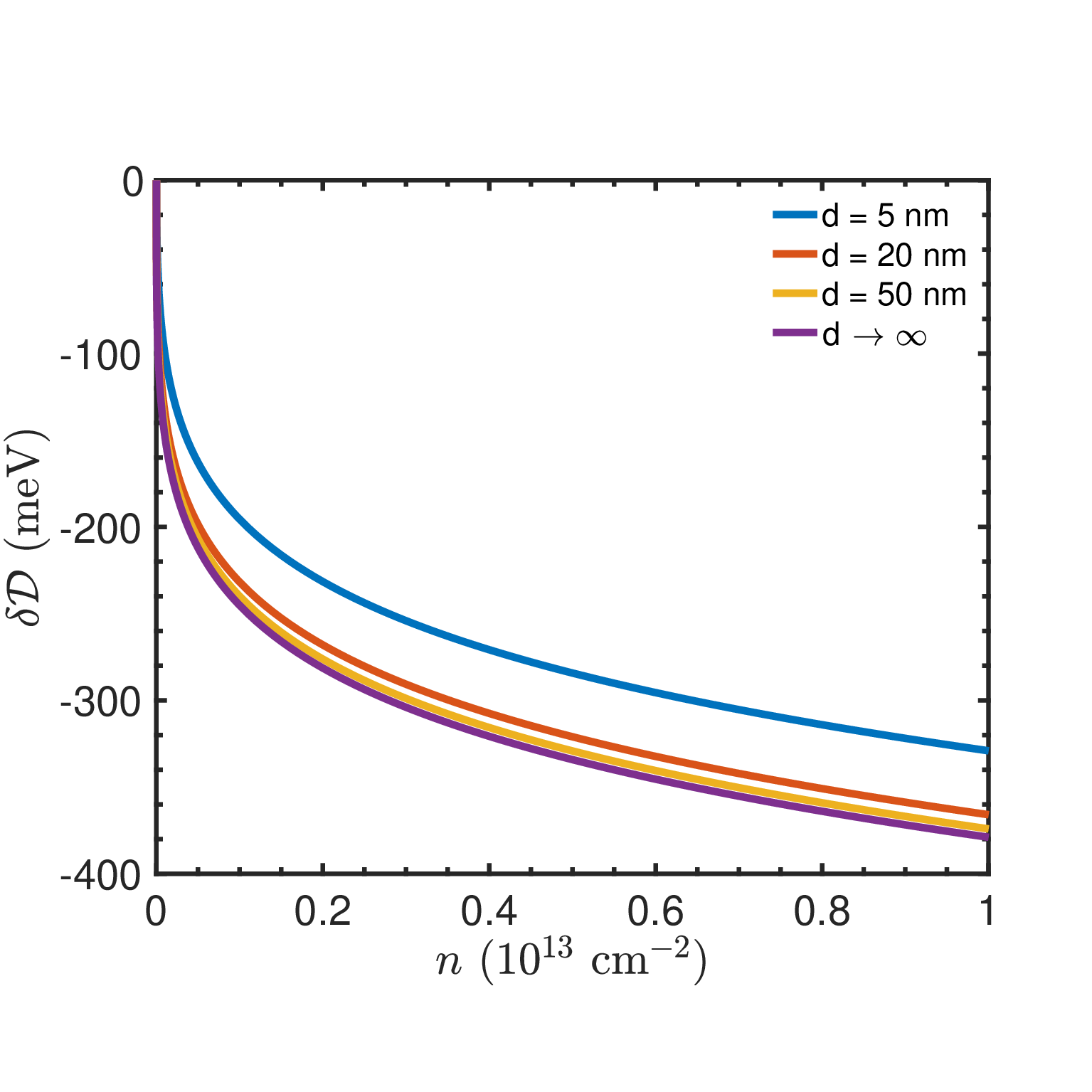}
		\vspace{-15pt}
	\caption{Band gap reduction due to conduction electrons for a MoS$_2$ on SiO$_2$ substrate device with a metallic gate located at distance $d$. Note that $\delta{\cal D}$ is calculated with respect to the insulator at each $d$, that is to say the correction to the insulating band gap is not included here. The effect of gate screening on band gap reduction is dramatic for low $d$, but by $d=50$ nm it is practically the same as the infinitely far gate limit and gate screening may be ignored.}
	\label{fig:MoS2_d}
\end{figure}
The full RPA screening calculations of band gap renormalization performed in this manuscript require careful application of complex analysis techniques. Simplified models have previously been used to account for screening of the Coulomb interaction. In  Appendix ~\ref{sec:compare} we compare our RPA results with a simple static screening model and a plasmon pole model.

\section{Comparison With Experimental Data} \label{sec:experiment}
The intrinsic band gap of 2D TMDs is large compared to the Fermi energy $\epsilon_F\ll\cal D$$_0\sim2eV$ to the extent that the dispersion may be approximated as truly parabolic. Thus when comparing our theory to experimental data $\cal D$$_0$ may always be tuned independently of $\delta{\cal D}$. 

We first consider MoS$_2$ on a SiO$_2$ substrate, with relevant material parameters given in Table~\ref{table:1}. Renormalized band gap as a function of n doping calculated from Eq~\ref{gapv} is plotted in Fig.~\ref{fig:MoS2}. The theoretical prediction is compared with available experimental data from Ref.~\cite{liu_direct_2019} and Ref.~\cite{yao_optically_2017}, where the latter is offset in intrinsic band gap ${\cal D}_0$ to allow for better comparison of $\delta{\cal D}$. For the theory curve ${\cal D}_0$ is set to ${\cal D}_0=2.43$ eV to best fit the data. It is also unclear if there is any initial doping in any of the systems due to impurities. Our theory reasonably agrees with the experimental $\delta{\cal D}$ from Ref.~\cite{liu_direct_2019} and is in excellent agreement with Ref.~\cite{yao_optically_2017}, which captures the dramatic low density reduction. 

In both Ref.~\cite{liu_direct_2019} and Ref.~\cite{yao_optically_2017} the gate is located a distance $d=285$ nm from the 2DEG. Hence according  to Eq.~\ref{igr} the correction $\delta{\cal D}_0$
  is very small, $\delta{\cal D}_0\approx 1$meV,
 and also $\delta{\cal D}$ can be calculated with $d\to\infty$, see 
  Fig.~\ref{fig:MoS2_d}.
Thus in the theoretical calculations presented in Fig.~\ref{fig:MoS2} the
effects of gate screening are negligible.

\begin{figure} [t]
	\centering
		\vspace{-20pt}
	\includegraphics[width=0.9\linewidth]{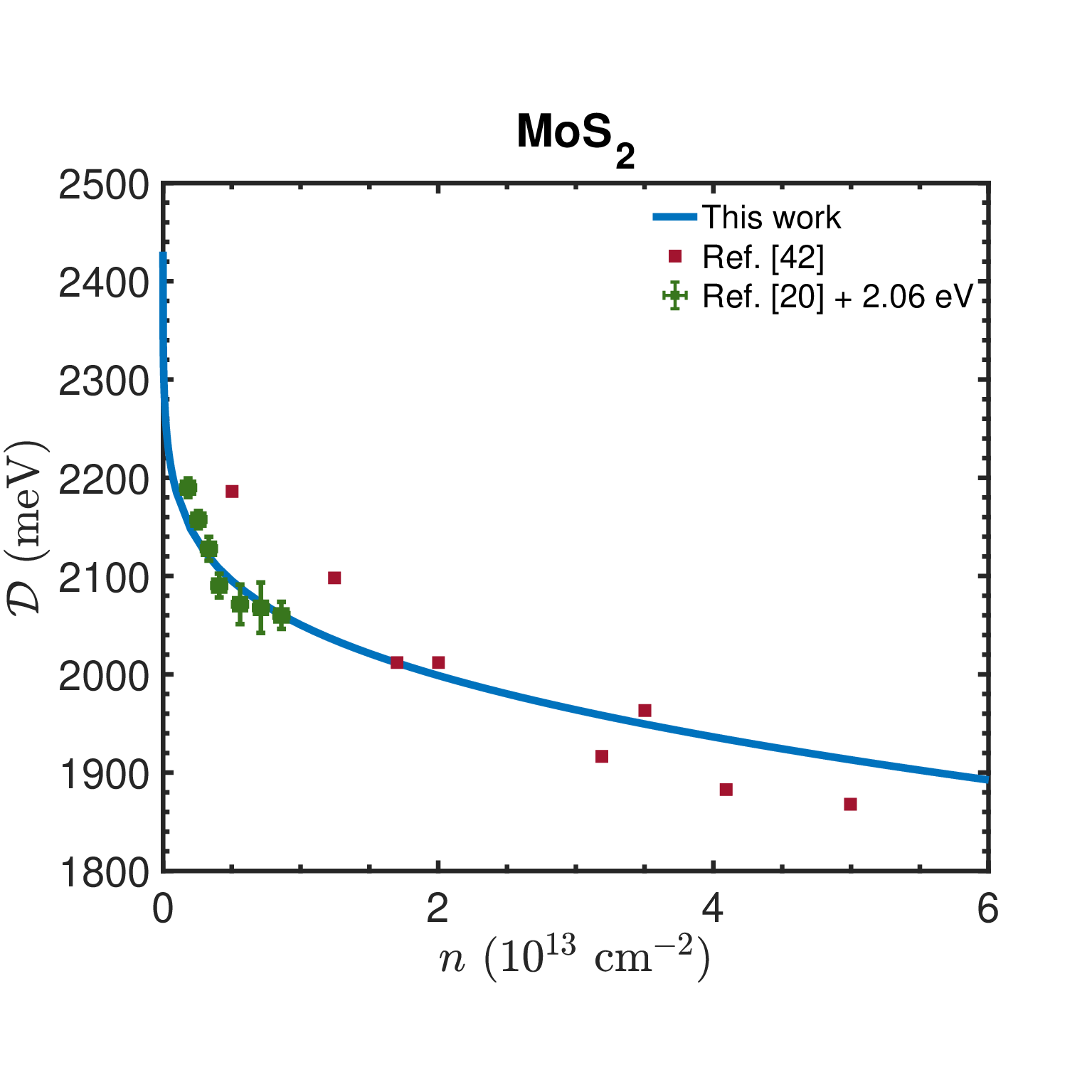}
		\vspace{-15pt}
	\caption{MoS$_2$ on SiO$_2$ substrate band gap renormalization calculation and experimental data from Ref.~\cite{liu_direct_2019} and Ref.~\cite{yao_optically_2017}. The $\delta{\cal D}$ data from Ref.~\cite{yao_optically_2017} has been shifted in $\cal D$ by $2.06$ eV for visualisation on the same axes (We are only concerned with renormalization, not absolute band gap). For the theoretical calculations $m_e=0.44$ m$_0$, $m_h=0.54$ m$_0$, $\varepsilon = 2.5$ and $r_0 = 3.4$ nm/$\varepsilon$. $\cal D$$_0$ is tuned to $2.43$ eV.}
	\label{fig:MoS2}
\end{figure}

\begin{table}[bp]
	\caption{Material parameters for the WSe$_2$ device.}
	\label{table:2}
	\begin{tabular}{lll}
		\hline
		\multicolumn{3}{l}{WSe$_2$. Strata: hBN, air} \\ \hline
		Parameter & Value & Description  \\ \hline
		$m_e$ & $0.29$ m$_0$ \cite{kormanyos_k_2015} & Effective electron mass \\
		$m_h$ & $0.42$ m$_0$ \cite{nguyen_visualizing_2019} & Effective hole mass \\
		$\varepsilon$ & $2.5$  \cite{nguyen_visualizing_2019,engdahl_excitons_2024} & Effective dielectric constant \\
		$r_0$ & $4.5$ nm/$\varepsilon$ \cite{engdahl_excitons_2024} & Characteristic screening length 
		\\ \hline
	\end{tabular}
\end{table}

\begin{figure} [t]
    \centering
    	\vspace{-20pt}
\includegraphics[width=0.9\linewidth]{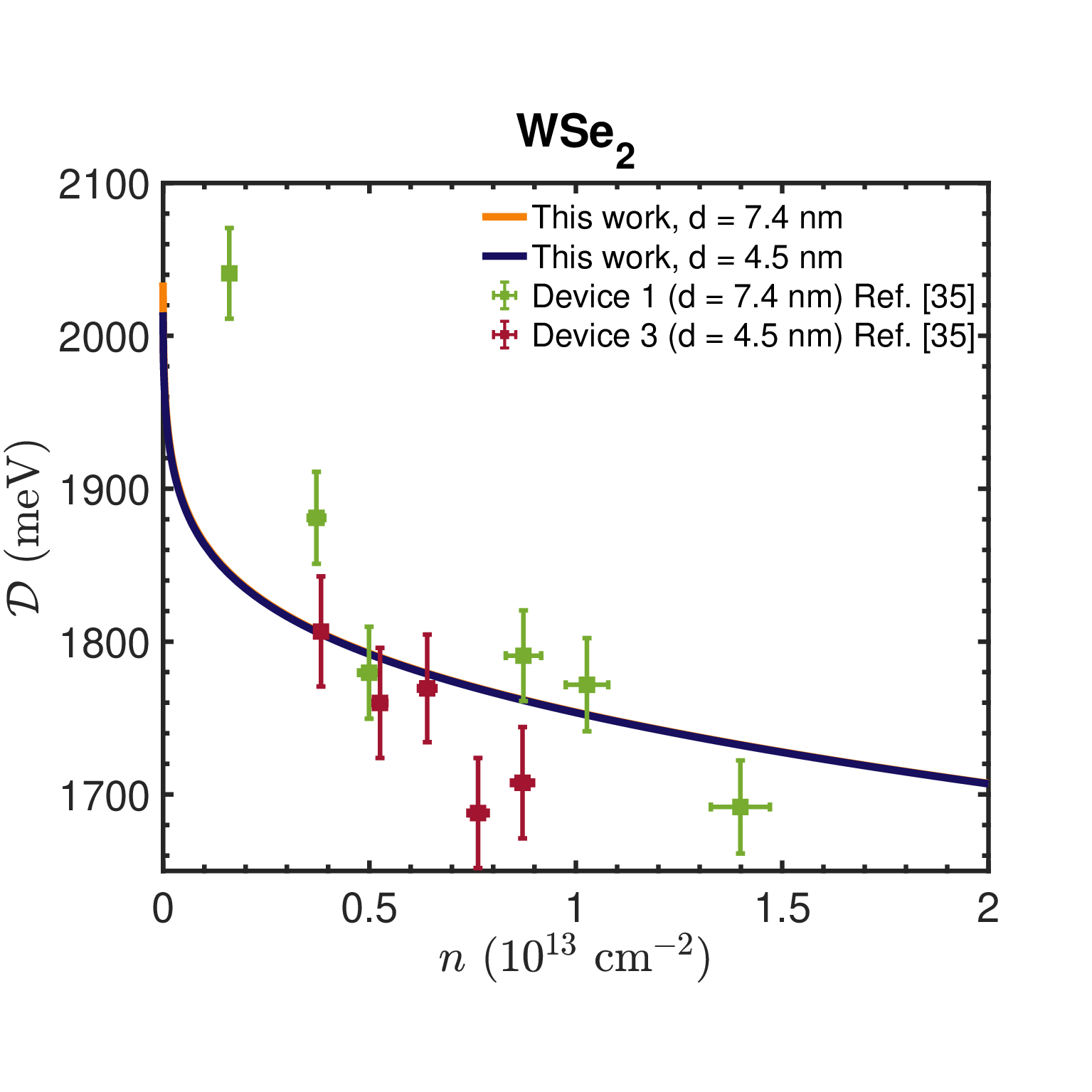}
	\vspace{-15pt}
\caption{WSe$_2$ on hBN substrate band gap renormalization calculations for the dynamical screening model and experimental data from Ref.~\cite{nguyen_visualizing_2019}. For the theoretical calculation $m_e=0.29$ m$_0$, $m_h=0.42$ m$_0$, $\varepsilon = 2.5$ and $r_0 = 4.5$ nm/$\varepsilon$. ${\cal D}_0$ is tuned to $2070$ meV and ${\cal D}={\cal D}_0+\delta{\cal D}_0+\delta{\cal D}$. Relative to $d\to\infty$, $\delta{\cal D}_0=-35.1$ meV at $d=7.4$ nm and $\delta{\cal D}_0=-54.6$ meV at $d=4.5$ nm.}
\label{fig:WSe2}
\end{figure}

While gate screening effects are negligible in the MoS$_2$ device presented in Fig.~\ref{fig:MoS2}, the devices from Ref.~\cite{nguyen_visualizing_2019} considered in Fig.~\ref{fig:WSe2} feature graphite gates at a distance $d<10$ nm from the 2DEG. In this case the effects of gate screening must be accounted for in both the calculation of band gap renormalization as a function of n doping and the calculation of insulator band gap. In Fig.~\ref{fig:WSe2} we present theoretical calculations of $\Delta$ vs $n$ with a gate located at a distance $d=7.4$ nm and  $d=4.5$ nm from the 2DEG and superimpose experimental data from Ref.~\cite{nguyen_visualizing_2019}. Parameters used in theoretical calculations for the WSe$_2$ device are presented in Table.~\ref{table:2}. As the 2DEG and strata are common across the two device we assume the insulating band gap in the $d\to\infty$ limit is the same in both devices, tuning ${\cal D}_0=2.07$ eV. The correction to the insulating band gap is easily calculated with Eq.~\ref{igr}, where $\delta{\cal D}_0=-35.1$ meV at $d=7.4$ nm and $\delta{\cal D}_0=-54.6$ meV at $d=4.5$ nm. 

Our theory is in reasonable agreement with the data from Ref.~\cite{nguyen_visualizing_2019}, although there is some discrepancy between theory and experimental measurements at low density. It must be noted that in this work we assume the system is free from impurities at $n=0$ and do not account for defects, whereas in reality there may be some level of intrinsic doping, where it is well known within the literature that TMD samples are high in impurities and defects \cite{mueller_exciton_2018,chaves_bandgap_2020}. Still, the large band gap reduction of $\sim300$ meV over the range $0<n<1.5\times10^{13}$ cm$^{-2}$ agrees well with the experimental data.

While this manuscript specifically discusses the role of screening from conduction electrons and a metallic gate in band gap renormalization, we point out that a significant renormalization of the band gap may also be caused by variation in dielectric environment. A variation of $\varepsilon$ has experimentally revealed band gap renormalization of around 250 meV for MoSe$_2$ between two different substrates \cite{zhang_bandgap_2016}. For comparison, consider the MoS$_2$ device from Table~\ref{table:1} at constant density $n=10^{12}$ cm$^{-2}$ and $d\to\infty$. At this density $\delta{\cal D}$ calculated from Eq.~\ref{gapv} is strongly dependent on $\varepsilon$, with a suspended sample ($\varepsilon=1.0$) having a band gap renormalization larger than a sample on a SiO$_2$ substrate ($\varepsilon=2.5$) by $200$ meV.  

\section{Summary} \label{sec:summary}
We have developed the theory for bandgap renormalization in 2D semiconductors using the Feynman diagrammatic technique traditionally used in QED. In our approach we consider full RPA screening and convert the integral to the imaginary frequency regime which is both analytically and numerically convenient. We derive expressions for band gap renormalization of the insulator due to gate screening, relative to an infinitely far gate, and of the metal due to both screening from conduction electrons and gate screening. Applying our methodology to 2D TMDs, we show the band gap significantly decreases at low density, scaling as $n^{1/3}$. We compare our theoretical predictions with available experimental data for MoS$_2$ and WSe$_2$ and find that our theory is in good agreement with experimental data. This dynamical screening self energy method may be modified and applied to systems with indirect band gaps or systems with a dispersion other than simple parabolic. 

\section{Acknowledgements} 
We acknowledge discussions with Ayd\i n Cem Keser, Zeb Krix, Iolanda Di Bernardo and Shaffique Adam. This work was supported by the Australian
Research Council Centre of Excellence in Future Low-
Energy Electronics Technologies (CE170100039). 

\appendix

\section{Comparison with approximations} \label{sec:compare}
Rather than consider the full dynamical screening from RPA, the screened interaction is often approximated with various models with differing levels of success. In this appendix we compare our theory to a static screening model and a plasmon pole model. Note that for simplicity we present solutions for the case $d\to\infty$.

In the static screening model, rather than considering dynamical screening from the full RPA chain, dynamical effects are ignored and a single value of frequency $\omega=0$ is used in the determination of the polarization operator in Eq.~\ref{ser}. The interaction no longer has cuts in the complex $\omega$ plane. In this regime the $\omega$ integration is readily solved via simple application of Cauchy's residue theorem. The total correction to the band gap for the case of static screening is given below in Eq.~\ref{eq:static}.

\begin{align} \label{eq:static}
	\delta{\cal D}_s=&- \frac{e^2}{\varepsilon r_0}ln(1+r_0p_F)+ \int_{p_F}^\infty \left[V^{(m)}_q(0)-
	V^{(ins)}_q\right] q \frac{dq}{2\pi}
\end{align}

Plasmon-pole models are commonly used theoretical models to take dynamical screening into account in the calculation of the quasiparticle band gap in 2D materials\cite{gao_dynamical_2016,gao_renormalization_2017,liang_carrier_2015,qiu_screening_2016}. The plasmon pole model involves placing all dynamical screening effects from the dynamical frequency polarization operator on the plasmon-pole and renormalizing the strength of the pole. It is easy to consider the self energy calculation in real frequency and see that plasmon effects dominate the self energy renormalization, thus validating the assumption of the plasmon pole description. We begin by rewriting the difference between the metallic and insulator interaction, where $\Omega_p$ is the plasmon pole strength and $\omega_p$ is the plasmon frequency which is obtained from a frequency dependent calculation of the poles of the interaction \ref{Vm}.

\begin{equation}\label{eq:V-ppm}
	\left[V^{(m)}_q(\omega)-
	V^{(ins)}_q\right] = \frac{1}{q}\left(\frac{\Omega_p^2(q)}{\omega^2-\omega_p^2(q)+i0}\right)
\end{equation}

The plasmon pole strength is weighted by the static screening dielectric function.

\begin{figure} [t]
	\centering
	\vspace{-20pt}
	\includegraphics[width=0.9\linewidth]{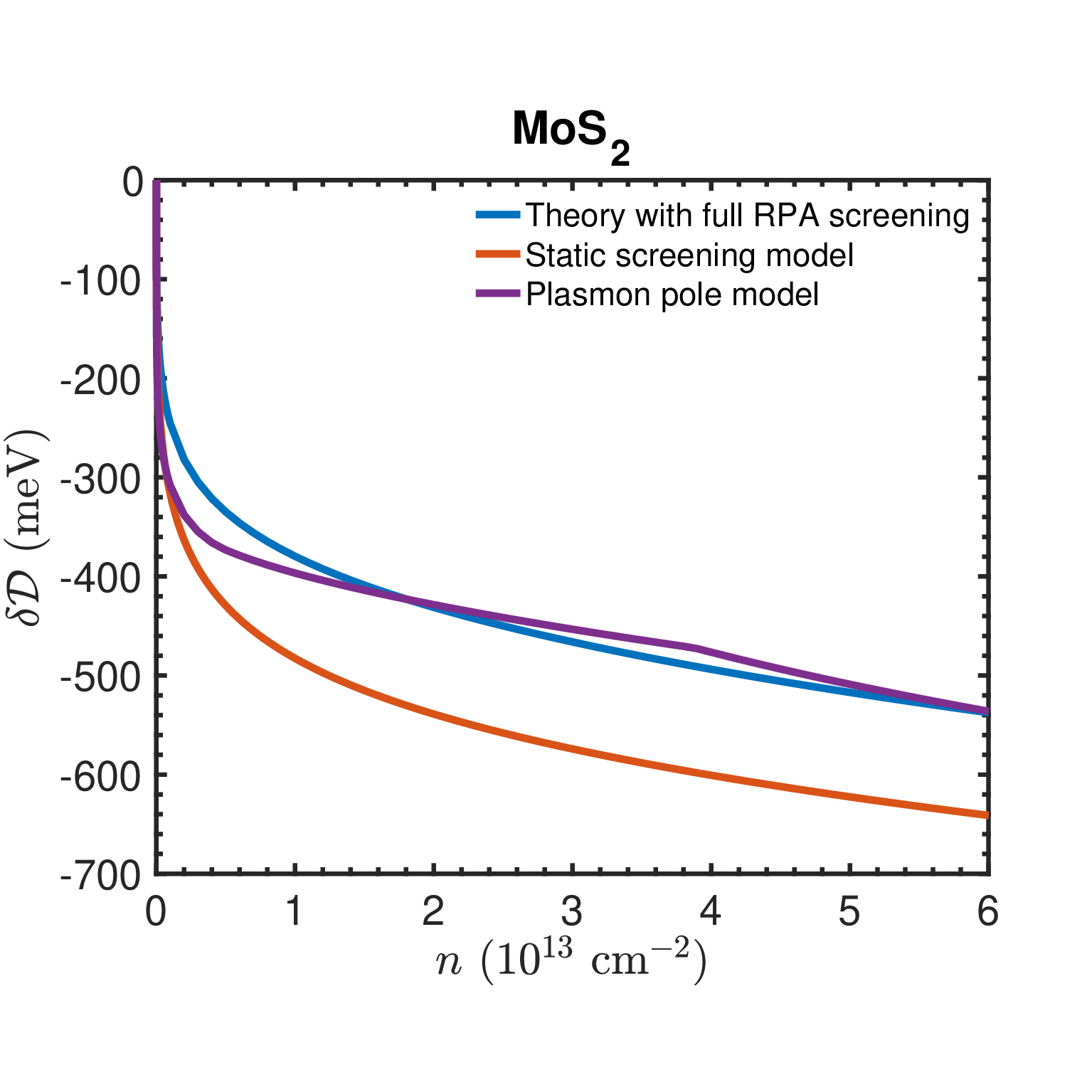}
	\vspace{-15pt}
	\caption{MoS$_2$ on SiO$_2$ substrate band gap renormalization calculations for three different screening models. The blue line is the dynamical screening model primarily considered in this work, the red line is a static screening model and the purple line is a plasmon pole model. Relative to the dynamical screening model, the static screening model overpredicts $\delta{\cal D}$ at all densities, whereas the plasmon pole model overpredicts $\delta{\cal D}$ at low density $n<10^{13}$ cm$^{-2}$ but agrees well with the full RPA calculation otherwise.}
	\label{fig:model_comparison}
\end{figure}

\begin{align}
	\Omega^2_p(q) =-q\omega_p^2(q)\left[V^{(m)}_q(0)-
	V^{(ins)}_q\right]
\end{align}

Substituting \ref{eq:V-ppm} into Eq.~\ref{ser}, the cuts of the interaction in the complex plane are replaced by simple poles. Finally, through application of Cauchy's residue theorem in the complex $\omega$ plane, one may derive an expression for total band gap renormalization in the plasmon pole approximation.

\begin{align}\label{eq:ppm}
	\delta{\cal D}_p &= -\frac{e^2}{\varepsilon r_0}ln(1+r_0p_F) + \int_{p_F}^\infty \frac{\Omega_p^2}{(\epsilon_q^{(e)})^2-\omega_p^2}\frac{dq}{2\pi} \nonumber\\ &+  \int_0^\infty \frac{\Omega_p^2 \left((\epsilon_q^{(e)})^2+(\epsilon_q^{(h)})^2\right)}{4\omega_p^2\left(-\omega_p-(\epsilon_q^{(e)})^2\right)\left(-\omega_p+(\epsilon_q^{(h)})^2\right)}\frac{dq}{2\pi}
\end{align}

Band gap reduction calculated with the exact expression Eq.~\ref{gapv}, the static screening approximation Eq.~\ref{eq:static} and the plasmon pole approximation Eq.~\ref{eq:ppm} are presented in Fig.~\ref{fig:model_comparison}, where the material parameters are taken from Table.~\ref{table:1} and the gate distance is $d\to\infty$. The plasmon pole model overpredicts $\delta{\cal D}$ for low density $n<10^{13}$ cm$^{-2}$ but is in good agreement with the exact calculation otherwise. Compared with the exact calculation, the static screening model overpredicts $\delta{\cal D}$ by $100-200$ meV at moderate to high carrier density. The cusp in the plasmon pole model is due to the second term in Eq.~\eqref{eq:ppm} vanishing, which occurs when the plasmon pole terminates at $q<p_F$, ie the plasmon pole exists entirely within the continuum. 

\begin{figure} [t]
	\centering
		\vspace{-20pt}
	\includegraphics[width=0.9\linewidth]{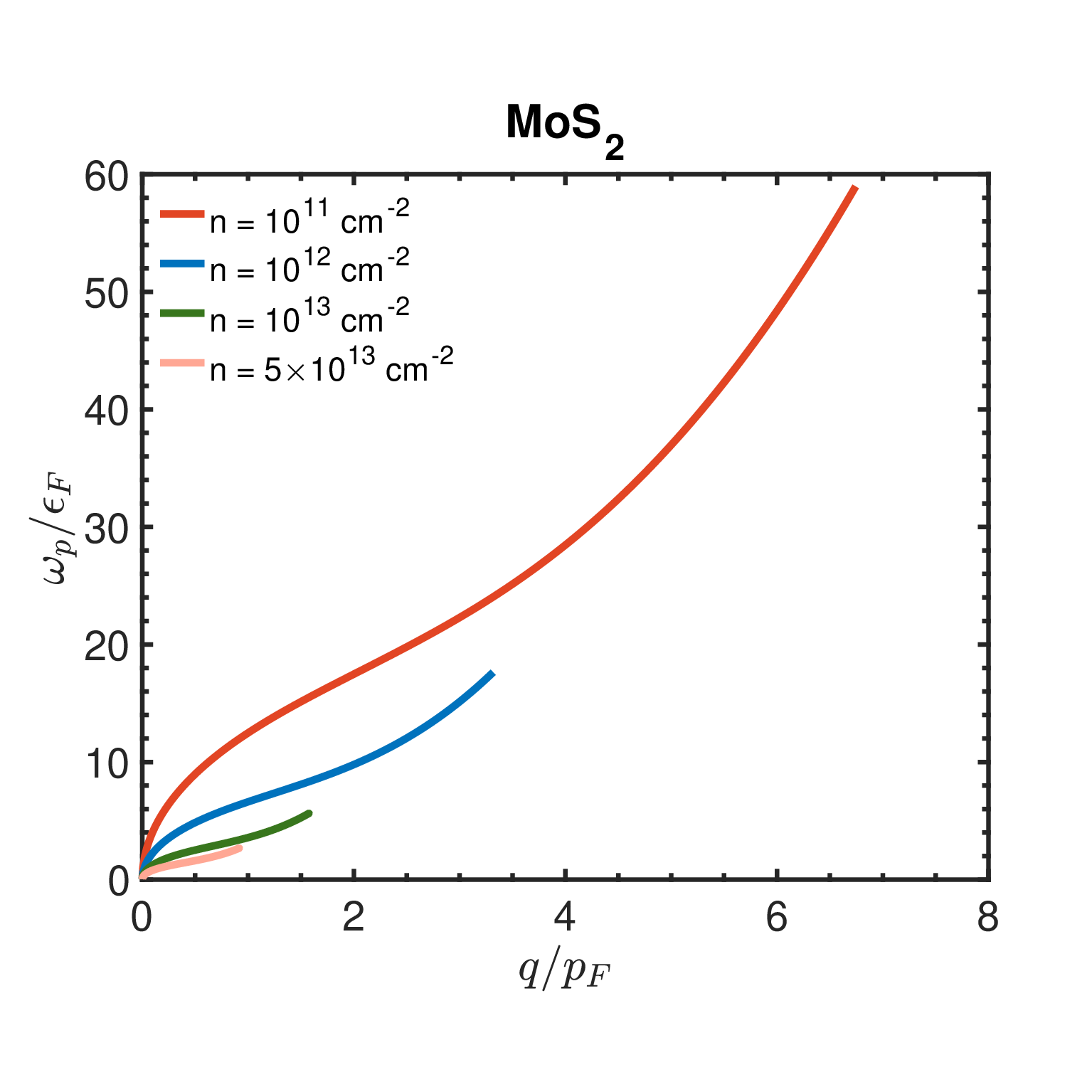}
		\vspace{-15pt}
	\caption{Plasmon frequency $\omega_p$ as a function of momentum $q$ for MoS$_2$ for a number of different carrier densities. Plasmon frequency is larger and plasmons exist up to larger momentum at lower densities. For this device $\omega_p$ terminates at $q=p_F$ at $n=3.9\times10^{13}$ cm$^{-2}$, dissolving into the continuum below this.}
	\label{fig:plasmons}%
\end{figure}

The plasmon pole model used in Fig.~\ref{fig:model_comparison} uses plasmon frequency $\omega_p$ that is obtained from a frequency dependent calculation of the RPA screened interaction with the exact plasmon frequency plotted in Fig.~\ref{fig:plasmons} as a function of momentum for a number of carrier densities. Clearly plasmons exist up to higher momenta at lower density, therefore contributing more to $\delta{\cal D}$ at low density than at higher density. For the MoS$_2$ device considered, the plasmon frequency  exists at momenta $q>p_F$ until $n=3.9\times10^{13}$ cm$^{-2}$. 

\begin{figure} [H]
	\centering
	\vspace{-20pt}
	\includegraphics[width=0.9\linewidth]{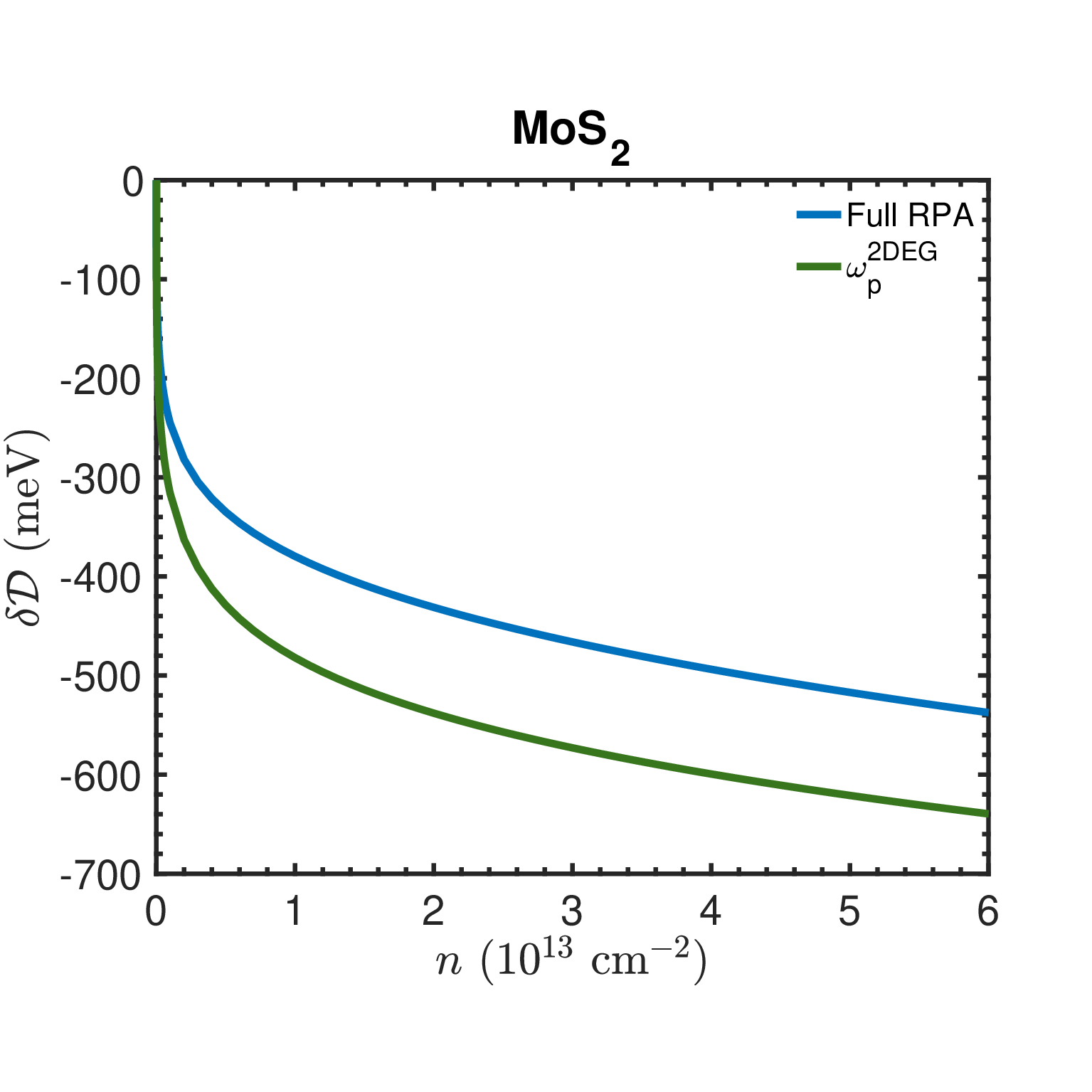}
	\vspace{-15pt}
	\caption{Comparison of $\delta{\cal D}$ for MoS$_2$ for full RPA screening and the plasmon pole model using $\omega_p^{2DEG}$, the plasmon frequency for a 2DEG within the RPA. $\omega_p^{2DEG}$ differs from the exact RPA $\omega_p$ for the 2D TMD as it does not account for the Rytova-Keldysh potential. Additionally, as this simplified plasmon pole model does not involve a frequency dependent calculation of the RPA screening for 2D TMD it does not include termination of plasmons. Hence the plasmon pole model overpredicts  $\delta{\cal D}$ compared with the full RPA calculation at every density.}
	\label{fig:ppm_compare}
\end{figure}

However, the procedure of first exactly determining plasmon frequency is more computationally intensive than computation of the exact calculation in imaginary frequency and for the case of simple parabolic band structure there is little justification for its use. Hence we consider the plasmon pole model where the 2DEG RPA plasmon dispersion relation\cite{czachor_dynamical_1982}

\begin{equation}
	\omega_p^{2DEG} = \sqrt{\frac{2\pi n q}{m}\left(1+\frac{q}{2}\right)^2\left(1+\frac{q^3}{8\pi n}+\frac{q^4}{32\pi n}\right)/\left(1+\frac{q}{4}\right)} . 
\end{equation}

This allows one to forego the frequency dependent calculation of the RPA screened interaction, however in doing so one neglects the calculation of plasmon termination. Thus this simplified model assumes plasmons exist to infinite momenta. We compare the results of the full RPA calculation with the plasmon pole model using $\omega_p^{2DEG}$ for Mos$_2$ in Fig.~\ref{fig:ppm_compare}. In this case the plasmon pole model overpredicts $\delta{\cal D}$ at all $n$, and in fact exactly reproduces the static screening model of Fig.~\ref{fig:model_comparison}. Thus plasmon termination is an important feature of a plasmon pole model that cannot be neglected. In this case there is little justification to using a plasmon pole model over the imaginary frequency full RPA calculation.
\bibliography{refs}
\end{document}